\documentstyle[12pt,sergio,epsfig]{article}

\def\nostrocostrutto#1\over#2{\mathrel{\mathop{\kern 0pt \rlap 
  {\raise.2ex\hbox{$#1$}}}
  \lower.9ex\hbox{\kern-.190em $#2$}}}
\def\gsim{\nostrocostrutto > \over \sim}   

\newcommand{\be}{\begin{equation}}
\newcommand{\ee}{\end{equation}}
\newcommand{\ba}{\begin{eqnarray}}
\newcommand{\ea}{\end{eqnarray}}

\newcommand{\eref}[1]{(\ref{#1})}      

\catcode`@=11
\newcount\@tempcntc
\def\@citex[#1]#2{\if@filesw\immediate\write\@auxout{\string\citation{#2}}\fi
  \@tempcnta\z@\@tempcntb\m@ne\def\@citea{}\@cite{\@for\@citeb:=#2\do
    {\@ifundefined
       {b@\@citeb}{\@citeo\@tempcntb\m@ne\@citea\def\@citea{,}{\bf ?}\@warning
       {Citation `\@citeb' on page \thepage \space undefined}}%
    {\setbox\z@\hbox{\global\@tempcntc0\csname b@\@citeb\endcsname\relax}%
     \ifnum\@tempcntc=\z@ \@citeo\@tempcntb\m@ne
       \@citea\def\@citea{,}\hbox{\csname b@\@citeb\endcsname}%
     \else
      \advance\@tempcntb\@ne
      \ifnum\@tempcntb=\@tempcntc
      \else\advance\@tempcntb\m@ne\@citeo
      \@tempcnta\@tempcntc\@tempcntb\@tempcntc\fi\fi}}\@citeo}{#1}}
\def\@citeo{\ifnum\@tempcnta>\@tempcntb\else\@citea\def\@citea{,}%
  \ifnum\@tempcnta=\@tempcntb\the\@tempcnta\else
   {\advance\@tempcnta\@ne\ifnum\@tempcnta=\@tempcntb \else \def\@citea{--}\fi
    \advance\@tempcnta\m@ne\the\@tempcnta\@citea\the\@tempcntb}\fi\fi}
\catcode`@=12

\begin{document}

\setcounter{page}{0}
\thispagestyle{empty}

\noindent
\rightline{Durham DTP/96/90}
\rightline{MPI-PhT/96-92}
\rightline{September 28th, 1996}
\vspace{1.0cm}
\begin{center}
{\Large \bf QCD coherence and} 
\end{center}
 \begin{center}
{\Large\bf the soft limit of the energy spectrum} 
\end{center}
\vspace{0.4cm} 
\begin{center}
VALERY A. KHOZE~$^{a, b, }$\footnote{e-mail: v.a.khoze@durham.ac.uk} \ , \ 
SERGIO LUPIA~$^{c, }$\footnote{e-mail: lupia@mppmu.mpg.de} \ , \ 
WOLFGANG OCHS~$^{c, }$\footnote{e-mail: wwo@mppmu.mpg.de} 
\end{center}

\begin{center}
$^a$ \ {\it Department of Physics\\
University of Durham, Durham DH1 3LE, UK}\\ 
\mbox{ }\\
$^b$ \ {\it Institute for Nuclear Physics\\
St. Petersburg, Gatchina, 188350, Russia}\\
\mbox{ }\\
$^c$ \ {\it Max-Planck-Institut f\"ur Physik \\
(Werner-Heisenberg-Institut) \\
F\"ohringer Ring 6, D-80805 Munich, Germany} 
\end{center}
\vspace{1.0cm}

\begin{abstract}
The analytical perturbative approach, if taken  to the 
limit of its applicability, allows one to predict an energy independent 
limit for 
the  one-particle invariant  density in QCD jets  $E \frac{dn}{d^3p}$ 
at very small momenta $p$. This is a direct consequence of the colour coherence
in soft gluon branching. The existing data on the charged and identified
particle inclusive spectra follow this prediction surprisingly well. 
Further tests of the perturbatively based picture in the soft region 
are discussed. 
\end{abstract}

\newpage

\section{Introduction}

One of the most striking predictions of the perturbative approach to QCD jet
physics \cite{bcm,dkmt1} 
is the depletion of the soft particle
production and the resulting approximately Gaussian shape of the inclusive
distribution in the variable $\xi = \log E_{jet}/E$ for particles
with energy $E$ in a jet of energy $E_{jet}$ (the so-called ``hump-backed
plateau'')\cite{dfk1,bcmm}. 
Due to the coherence of the gluon radiation it is not the softest partons but
those with intermediate energies ($E \sim E_{jet}^{0.3-0.4}$) which 
multiply most effectively in QCD cascades. We recall that the gluons 
of  long wave length are emitted by
the total colour current which is independent of the internal structure of the
jet and is conserved when the partons split. Applying the hypothesis of the
Local Parton Hadron duality (LPHD)\cite{adkt1}, one then comes to the
conclusion that the hadron spectrum at low momentum $p$ should be 
nearly independent of
the jet energy $E_{jet}$\cite{adkt1}. 
 
It is a very important question to understand whether there is a smooth
transition between the purely perturbative regime and the soft momentum region.
It looks quite intriguing that the shape of the measured charged particle
energy spectra in $e^+e^-$ annihilation appears to be surprisingly
close over the whole momentum range (down to small momenta of a few hundred
MeV) to the perturbative predictions based on the Modified Leading Log
Approximation (MLLA)\cite{dkmt1,dt,ahm1,ko}. Moreover, the momentum spectra in
the variable $\log p$, 
 measured by the  TASSO \cite{tasso} and 
 OPAL Collaborations \cite{opal1} provided first 
 evidence in favour of the scaling
 behaviour at small momenta: as discussed in \cite{vakcar} these data
 could be considered as a confirmation of the basic ideas of QCD coherence and
 LPHD. 
A further confirmation of this picture has been provided recently in the
study of the invariant particle density  $E dn/d^3p$ 
at low momenta and also in the low $cms$ energy region. 
The spectra are found to be in a good agreement with
the scaling behaviour and
with the analytical perturbative expectations which become sensitive
to the running of the coupling $\alpha_s$ at small scales 
\cite{lo}. 

In this paper we focus on the soft limit of the parton spectra, 
and how its scaling behaviour arises in the analytical
calculations for the 
coherent parton cascade. 
We consider the invaraint density $E dn/d^3p \equiv dn/dy d^2p_T$ in the limit
of vanishing rapidity $y$ and transverse momentum $p_T$ or, equivalently, for
vanishing momentum $|\vec p | \equiv p$, i.e., 
\be
I_0 = \lim_{y \to 0, p_T \to 0} E \frac{dn}{d^3p} = 
\frac{1}{2} \lim_{p \to 0} E \frac{dn}{d^3p} 
\label{izero}
\ee
where the factor 1/2 takes into account that both hemisphere are added in the
limit $p \to$ 0. 
If the dual description of hadronic and partonic final states is really
adequate down to very small momenta, the finite, energy independent limit of
the invariant hadronic density $I_0$ is expected. 

This will be examined  for both charged and  identified hadrons
in the full energy range explored so far in $e^+e^-$ annihilation.
A possible growth of the spectrum
in the soft region could indicate that either coherence or the local duality
(or both) breaks down. Since colour coherence is a general property of QCD as a
gauge theory, it is the LPHD concept that is tested in the measurements of the
soft hadron distributions.
 Therefore, such studies could provide one with 
additional information on the physics of confinement.

The scaling properties of the invariant cross section or density
have been discussed in the early seventies in the context of the parton
model \cite{bj} and Feynman-scaling \cite{feynman} (see e.g. \cite{perl}).
Note that scaling in the parton model refers to the $p_T$-integrated spectra.
In the QCD bremsstrahlung picture of multiparticle production discussed here 
the $p_T$-integrated central
rapidity plateau rises with $cms$ energy, and scaling behaviour 
is retained only in the large wavelength limit (\ref{izero}). 

It is  important to  test further the QCD origin of this scaling
behaviour. This can be accomplished by studying the soft radiation from
the emitters (``parton antennae'') corresponding to  
different effective colour charges.
The good scaling property of the density $I_0$ 
in $e^+e^-$ annihilation suggests taking this number as a standard scale in
the comparison with other processes. 
Some of the results of this paper and further details have also been
presented elsewhere \cite{conf}.

\section{Particle production in the soft limit in DLA and MLLA}

\noindent {\it DLA predictions.}\\
We consider first the analytical predictions for the energy spectrum of
partons near the soft limit. The asymptotic behaviour of the energy spectrum
is obtained from the Double Log Approximation (DLA) in which energy
conservation is neglected and only the leading singularities in the parton
splitting functions are kept. 
The evolution equation of the inclusive energy 
distribution of partons $p$
originating from a primary parton $A$ is given by\cite{dfk1}: 
\be
D_A^p(\xi,Y) = \delta_A^p\delta(\xi) + \int_0^{\xi} d\xi' \int_0^{Y-\xi} 
dy' \frac{C_A}{N_C} \gamma_0^2(y') D_g^p(\xi',y') \; . 
\label{evoleq}
\ee
Here  we have used the logarithmic variables 
 $\xi = \log (1/x) = \log (Q/E)$ and $Y = \log (Q/Q_0)$ with
$E$ the particle energy and $Q$ the jet virtuality ($Q=P\Theta$ for
a jet of primary momentum $P$ and half opening angle $\Theta$); 
 $C_A$ is the respective colour factor, i.e., $C_g = N_C$ and $C_q = C_F$;
 $\gamma_0$ denotes the anomalous dimension of multiplicity and 
is related to the 
QCD running coupling by $\gamma_0^2 = 4 N_C \alpha_s / 2 \pi$
or $\gamma_0^2 = \beta^2/\log(p_T/\Lambda)$ with $\beta^2=4 N_C /b$,
$b \equiv (11 N_c - 2 n_f)/3$;  
$\Lambda$ is the QCD-scale and  $N_C$ and $n_f$ are 
the number of colours and of flavours respectively. 
The shower evolution is cut off by 
$Q_0$, such that the transverse momentum $p_T \ge Q_0$. 

As can be seen from (\ref{evoleq}) the $\xi$-spectrum vanishes in the soft
limit $\xi\to Y$ ($E\to Q_0$). The behaviour near this limit can be found by
the iterative solution of (\ref{evoleq}). With two iterations one obtains
\be
D_A^g(\xi,Y) = \delta_A^g \delta(\xi) + \frac{C_A}{N_C} 
\beta^2 \log \left( 1 + \frac{Y-\xi}{\lambda}  \right) 
\left[ 1 + \frac{\beta^2 \int_0^{Y-\xi} 
d\tau \log (1 + \frac{\tau}{\lambda}) 
\log (1 + \frac{\xi}{\tau+\lambda})}{\log (1 + \frac{Y-\xi}{\lambda})} \right] 
+ \dots
\label{duetermini}
\ee
where $\lambda = \log Q_0/\Lambda$. 
The second term of order $\beta^2$ corresponds to the emission of a single gluon 
and  yields the leading contribution for $E \to
Q_0$. It is proportional to the colour charge factor of the primary parton.
Furthermore, this term does not depend on the $cms$ energy, contrary to
 the higher order terms which provide the rise 
of the spectrum for large $E$ with increasing $\sqrt{s}$. 
The spectrum vanishes at $E \to Q_0$ as
\be 
D_A^g(\xi,Y) \sim Y- \xi \sim  \log E/Q_0 \sim E-Q_0.
\label{limitth} 
\ee
An exact solution is known for fixed $\alpha_s$ \cite{dfk1}
with similar behaviour near
the boundaries as above in (\ref{duetermini}) and (\ref{limitth}).

\noindent {\it MLLA predictions.}\\
A considerable improvement in accuracy for nonasymptotic energies is
obtained in the application of the MLLA \cite{dt,ahm1,dkmt1}
which takes into account
systematically all the corrections of relative order $\sqrt{\alpha_s}$
based on the evolution equation which includes energy conservation
and the exact form of the parton splitting functions. The leading 
high energy MLLA contribution obeys a differential equation 
(see also \cite{dkmt1,lo}, with  $D(\xi, Y, \lambda) \equiv D_g^g$ )
\be
\frac{\partial^2 D(\xi, Y, \lambda)}{\partial \xi \partial Y} +
\frac{\partial^2 D(\xi,Y,\lambda)}{dY^2} -
\gamma_0^2 D(\xi,Y,\lambda) = 
- a \biggl(\frac{\partial}{\partial \xi} +  \frac{\partial}{\partial Y}\biggr) 
\biggl(\frac{\alpha_s(Y+\lambda)}{2\pi} D(\xi,Y,\lambda)\biggr)
\label{deq}
\ee
where $a=\frac{11}{3}N_C+\frac{2n_f}{3N_C^2}$. The limit $a=0$ corresponds
to the DLA as in (\ref{evoleq}). In case of fixed $\alpha_s$ the exact MLLA
solution can be obtained \cite{lo} (see also \cite{dt}) by multiplying the
DLA result 
with the factor
$\exp(- a\gamma_0^2(Y-\xi)/4N_C)$. Similarly, for running $\alpha_s$
one can make an ansatz
\begin{equation}
D(\xi,Y,\lambda) = D(\xi,Y,\lambda)|_{DLA} \exp \biggl[-a\int^Y_\xi
\gamma_0^2(y)/(4N_C) dy \biggr] 
\label{dmlla}
\end{equation}
which solves Eq. (\ref{deq}) approximately in the region around 
$\xi\sim Y$.\footnote{When eq.~(\ref{dmlla}) is inserted into eq.~(\ref{deq}) 
all terms cancel except for the small difference $(\partial D_{DLA}/\partial Y) 
\exp [-a\int^Y_\xi \gamma_0^2(y)/(4N_C) dy] [ \frac{a \alpha_s(\xi)}{2 \pi} - 
\frac{a \alpha_s(Y)}{2 \pi}]$, which vanishes at $\xi=Y$. Exponential
damping factors of this type 
are known to take into account single log corrections\cite{adk80}.}  
Note that the exponent in (\ref{dmlla}) 
can be written as 
$ - \frac{a}{b} 
       \log \biggl( \frac{Y+\lambda}{\xi+\lambda} \biggr) $
and behaves near the boundary
like $-\frac{a}{b} \frac{Y-\xi}{Y+\lambda}$,
so the slope of the distribution varies with energy like the coupling
$\alpha_s\sim\gamma_0^2= 4 N_C/[b (Y+\lambda)]$ contrary to the case with fixed
coupling.\\

\noindent {\it Phase-space effects in the soft region.}\\
While the limiting behaviour of the partonic energy spectrum in the soft
region follows from the general principle of colour coherence the detailed
form of the observable hadronic spectrum is predicted uniquely from the 
LPHD hypothesis only for $E \simeq p \gg Q_0$, but not  near the 
kinematical boundary because of the sensitivity of the spectrum 
to the cut-off procedure. In the MLLA the partons are
treated as massless with energy $E=p\geq p_T \geq Q_0$, so $\xi\leq Y$.
Experimental hadronic spectra are usually presented as function of momenta
$p$ or $\xi_p=\log (1/x_p)$ which is not limited from above. The same
kinematic limit for partons and hadrons is obtained if the hadronic mass
$m_h$ and the partonic $p_T$ cut-off $Q_0$ are taken the same.

For the relation between parton and hadron distributions one may require
that the invariant density $E dn/d^3p$ of hadrons approaches a constant limit
for $p\to 0$ as is observed experimentally. For the spectra which
vanish linearly as in (\ref{limitth}) this is achieved by relating 
the hadron and parton spectra in a single $A$-jet as \cite{dfk1,lo}
\be
E_h \frac{dn(\xi_E)}{dp_h} = K_h E_p \frac{dn(\xi_E)}{dp_p}
    \equiv K_h D_A^g(\xi_E,Y)
    \label{phrel}
\ee
with $E_h=\sqrt{p_h^2+Q_0^2}=E_p \equiv E
\geq Q_0$ and $\xi_E\equiv \xi=\log Q/E$, 
where $K_h$ is a normalization
parameter. If hadrons from both hemispheres are added, $K_h$ should be replaced
by $2K_h$. 
Then, indeed, for hadrons the invariant density
$E dn/d^3p = K_h D_A^g(\xi,Y)/4\pi (E_h^2-Q^2_0)$
approaches the finite limit as in (\ref{izero})
\be
I_0=K_h \frac{C_A\beta^2}{8\pi N_C\lambda Q_0^2}.
\label{limit}
\ee 
In the fixed $\alpha_s$ limit $\beta^2/\lambda$ is replaced by $\gamma_0$
and $I_0\sim 1/Q_0^2$.
With prescription (\ref{phrel}) the moments of the full energy spectrum 
$D(\xi,Y)$ are
well described by the MLLA formulae and with $Q_0=270$ MeV in a wide energy
range \cite{lo}.

The relation (\ref{phrel}) is not unique, however. We found that the
alternative prescription based on phase space arguments, 
$dn/d\xi_p \simeq (p/E)^3 D(\xi_E,Y)$  (see e.g.~\cite{DKTInt,dkt9}): 
\be
E_h\frac{dn}{d^3p_h} = K_h \biggl(\frac{1}{4 \pi E^2}\biggr) D_A^g(\xi_E,Y)
\label{phrel1}
\ee
works well in $e^+e^-$ annihilation 
for charged pions ($K_h = K_{\pi}$) 
in the whole energy region and for 
charged particles ($K_h = K_{ch}$) in the low energy region. In this
application, including both hemispheres, $D_A^g$ in eq.~\eref{phrel1} is replaced
by $ 2 \cdot 4/9 D_g^g$, using for $D_g^g$ the MLLA limiting spectrum 
\cite{adkt1,DKTInt} with
$Q_0=\Lambda=138$ MeV. The low cut-off mass is plausible in this region,
which is dominated by pions. 

\noindent {\it Behaviour of analytical results}\\
In order to illustrate the above analytical results we compare in Fig.~1
the various approximations (in case of running
$\alpha_s$) for low particle energies
using the relation (\ref{phrel}) between parton and hadron spectra:
the DLA result in the approximation (\ref{duetermini}) with and without
MLLA exponential factor (\ref{dmlla}) for $\lambda=0.01$ 
and also the ``limiting spectrum", which solves the MLLA equations
for $\lambda=0$ in the full kinematic region, except very close to 
the boundary $\xi=Y$ where it stays finite.
The normalization of the limiting spectrum is as in fits to $e^+e^-$
annihilation; the normalizations of the
DLA and MLLA curves are chosen to approach  the
limiting spectrum for energies $E~\geq~0.5$ GeV. 

These results show that in all cases considered, DLA and MLLA for both
fixed and running $\alpha_s$, the single particle invariant density
 approaches an
energy independent value in the soft limit $\xi\to Y$ ($E\to Q_0$). This
originates from the soft gluon emission contribution of order $\alpha_s$ 
which is determined
by the total colour charge of the primary partons due to the colour
coherence. In this limit the MLLA converges towards the DLA as the energy
conservation constraints are unimportant and the parton splitting functions
are only probed for very small fractional momenta.

\section{Discussion of experimental results}

\noindent{\it Inclusive charged  particle spectra}\\ 
Fig.~2 shows 
the charged particle  invariant density in $e^+e^-$ annihilation, $E dn/d^3p$, 
as a function of the particle energy $E$ at different $cms$ energies 
ranging from 3 GeV up to LEP-1.5 $cms$ 
energy (133 GeV)\cite{tasso,opal1,mark1,data}. 
The effective mass value of 
$Q_0$ = 270 MeV has been used in the kinematical relations. As shown in
\cite{lo} this allows a reasonably good description of the moments of the
energy spectra by the MLLA limiting formulae over a large energy interval. 
A fit using eq.~\eref{phrel1} is shown elsewhere (see last reference in 
\cite{conf}).

It is remarkable that the data from all  $cms$ energies tend to converge in the
soft limit: we find $2 I_0 \simeq$ 6-8 GeV$^{-2}$ (using $Q_0$ = 270 MeV) and 
$2 I_0 \simeq$ 4-6 GeV$^{-2}$ (using $Q_0$ = 138 MeV). 
Inspecting the soft limit more closely, the 
LEP data seem to tend to a limiting value larger by about 20\% as compared to
the lower energy data. This may be well due to 
the overall  systematic effect in the relative normalization 
of the different experiments. In this
respect, it is instructive to recall that both TASSO and  the LEP
Collaborations have collected their 
data at different $cms$ energies with the same
accelerators and detectors, thus avoiding within their data samples any
problems due to the relative normalization; 
in these samples the scaling works well.

A possible source of scaling violations are weak decays (for example kaons, heavy
quarks). Particles from such decays (from the new ``antenna'') would add
incoherently to the particles produced from the primary quarks and thereby
could yield a rise of the soft particle spectrum with increasing energy. It
would be interesting to disentangle these possibilities experimentally and to
find out the size of primary scaling violation, if any.

The data also compare well with the MLLA analytical predictions, not only
concerning the energy independence of the soft limit $I_0$ but also the
considerable energy dependence of the initial slope which is largely due to the
running of $\alpha_s$ in the MLLA damping factor (\ref{dmlla}).\\

\noindent{\it Identified particles' spectra}\\ 
Let us consider now the invariant density for $\pi$, $K$ and $p$. 
In this case, the simplest assumption is to identify with the mass scale 
the particle mass itself. 
Fig. 3 shows
the invariant cross section $E dn/d^3p$ 
as a function of the particle energy $E$ for 
charged pions, charged kaons and  protons, as derived from 
the  inclusive momentum spectra measured  at $cms$ energies from
1.6 GeV to 91 GeV\cite{data,datapion}. 
In all cases the data tend towards a common limit for $E \to m_h$ ($p \to 0$). 

Note that the scale for $I_0$ in eq.~\eref{limit} is given not by
$\Lambda^{-2}$, but by $Q_0^{-2}$, which 
can be related to an effective particle mass. 
Indeed, $I_0$ shows a clear mass dependence ($2 I_0^{\pi}$ = 5--8 GeV$^{-2}$, 
$2 I_0^{K}$ = 0.3--0.4 GeV$^{-2}$, $2 I_0^p$ = 0.08--0.10 GeV$^{-2}$). 
This is consistent with $I_0^h \simeq m_h^{-2}$ (after normalization to
$I_0^{\pi}$, one would predict 
 $I_0^{K}$ = 0.37--0.61 GeV$^{-2}$, $I_0^p$ = 0.10--0.17 GeV$^{-2}$). 
Inclusion of the logarithmic correction $1/\lambda$ in eq.~\eref{limit} would,
however, require rather small values of $\Lambda \simeq$ 100 MeV.
The measured values of $I_0^h$ are also in good agreement with the
expectations following from eq.~\eref{phrel1} and the finite limit at $E\to
Q_0$ of the limiting spectrum.
Alternatively, one can use in the fits effective scales $Q_0$ which deviate
from the respective hadron masses; this is also suggested by the 
 fits to the peak region of the $\xi$-distribution\footnote{
For example, the maximum $\xi^*$ of the distribution does not
follow the expectations ($\xi^* \simeq 1/2 \log E_{jet}/Q_0$ in
DLA) with $Q_0=m_h$\cite{delphimaximum}.}. 
This intriguing subject
certainly deserves further detailed studies.

\section{Further tests of the dual description of parton and hadron cascades 
in $e^+e^-$ annihilation.} 

It is suggestive to relate the observed behaviour of particle spectra
for vanishing momenta to the expected behaviour of soft gluon emission off the
primary partons. 
The particle rate $I_0$ in this limit (see eq.~\eref{limit}) 
cannot be directly predicted, as
it depends on the normalization factor and the cut-off parameter $Q_0$. 
However, the remarkable scaling behaviour of  $I_0$ 
in $e^+e^-$ annihilation provides one with a  standard scale for 
 the  comparison with other processes. 
 It would be very interesting to study whether the soft
 particle production is universal, i.e., a purely hadronic quantity  
 or whether the intensity $I_0$ indeed depends  
on the  colour topology  of the primary active partons in the collisions
process. This could provide one with 
a  direct support of the dual description of soft particle
production in terms of the QCD bremsstrahlung. 

In what follows we consider some possibilities to test this hypothesis further
and discuss what one may expect from the comparison of the $e^+e^-$ 
reaction with other collision processes.\\ 

\noindent{\it gg final state}\\
A direct test of the relevance of QCD coherence for the soft production limit is
the comparison of the $q\bar q$ with the $gg$ colour singlet final state. In
the $gg$ final state the same line of argument applies as above for the $q \bar
q$ final state, but the colour charge factor is increased by $N_C/C_F$ = 9/4 in
 eq.~\eref{limit} implying roughly a doubling of the soft 
 radiation intensity $I_0$.
This increased radiation has been suggested originally for the global
multiplicity in gluon jets\cite{bg}; in the real life experiments 
various effects keep the
observed increase considerably below this limit (see e.g. \cite{ko}). 
As the soft radiation is not strongly 
influenced by energy-momentum constraints (see the 
DLA-MLLA comparison above) the effects of the 
different colour charges could be more pronounced in this case. 

An approximate
realization of a colour octet antenna is possible in $e^+e^- \to q \bar q g$
with the gluon recoiling against a quasi--collinear $q \bar q$ pair\cite{gary} 
(for a recent study of such events, see \cite{opalqq}). 
The production rate of  soft particles in such events is
expected to be increased if it really is related to 
the effective colour charge of the primary partons.\\ 

\noindent{\it 3-jet events}\\ 
To be more quantitative, we consider the soft radiation in 3-jet $q\bar q g$
events of arbitrary jet orientation into a cone perpendicular to the production
plane and compare it to the radiation into the same cone in a 2-jet $q\bar q$
event again perpendicular to the primary $q\bar q$ directions. The analysis of
these configurations avoids the integration over the $k_T \ge Q_0$ boundaries
along the jets. 

We restrict ourselves to calculations of the soft gluon bremsstrahlung of
order $\alpha_s$. From the experience above with two-jet events this
contribution dominates in the soft limit whereas higher order contributions
take over with increasing particle energy. Calculations of this type have been
successfully applied to the string/drag phenomena\cite{lund,adkt} which refer
to the particle multiplicity flows projected onto the 3-jet production plane. 

Let us consider first the soft radiation into arbitrary direction $\vec{n}$
from a $q\bar q$ antenna pointing in directions $\vec{n}_i$ and
$\vec{n}_j$\cite{adkt}:
\be
dN_{q\bar q} = \frac{dp}{p} d\Omega_{\vec{n}} 
\frac{\alpha_s}{(2 \pi)^2} W^{q\bar q}(\vec{n}) \quad , \quad 
W^{q\bar q}(\vec{n}) = 2 C_F (\widehat{ij}) 
\label{wqq} 
\ee
with 
$(\widehat{ij}) = a_{ij}/( a_i a_j)$, $a_{ij} = (1 - \vec{n}_i
\vec{n}_j)$ and $a_i = (1 - \vec{n} 
\vec{n}_i)$. 
Such an antenna is realized, for example, in a $q \bar q \gamma$ event, and can
be obtained by the appropriate Lorentz boost from the $q\bar q$ rest frame.
The soft gluon radiation in a $q \bar q g$ event is given as in \eref{wqq} but
with the angular factor
\be
W^{q\bar q g}(\vec{n}) = N_C [ (\widehat{1+}) + (\widehat{1-}) 
- \frac{1}{N_C^2}  (\widehat{+-}) ] 
\ee
where $(+,-,1)$ refer to $(q,\bar q,g)$. 

For the radiation perpendicular to
the primary partons $(\widehat{ij}) = a_{ij} = 1 - \cos \Theta_{ij}$, with
relative angles $\Theta_{ij}$ between the primary partons $i$ and $j$. 
For 2-jet events of arbitrary orientation one obtains in this case 
\be
W^{q\bar q }_{\perp}(\Theta_{+-}) = 2 C_F ( 1 - \cos \Theta_{+-}) \;  .
\ee
Correspondingly the ratio $R_{\perp}$ of the soft particle yield in 3-jet events to
that of 2-jet events in their own rest frame 
($W_{\perp}^{q\bar q}(\pi) = 4 C_F$) is given by
\be
R_{\perp} \equiv \frac{dN_{\perp}^{q\bar q g}}{dN_{\perp}^{q\bar q}} =
\frac{N_C}{4 C_F} [ 2 - \cos \Theta_{1+} - \cos \Theta_{1-} - \frac{1}{N_C^2}
(1 - \cos \Theta_{+-} ) ] 
\label{rperp}
\ee
It is also interesting to note the difference of this prediction to the large
$N_C$ approximation in which the $q\bar q g$ event is treated as a
superposition of two $q\bar q$ dipoles (see, e.g., \cite{lund}). 
In this case the last term in eq.~\eref{rperp} drops out 
and $C_F = (N_C^2 - 1)/2 N_C \simeq N_C/2$. This yields 
\be
R_{\perp} \simeq 
\frac{1}{2} [ 2 - \cos \Theta_{1+} - \cos \Theta_{1-} ] \qquad \hbox{(large
$N_C$)}
\label{rperpln}
\ee
Predictions from these formulae for $R_{\perp}$ are presented in
Table~\eref{tableperp} for various relative angles $\Theta_{ij}$. Note, in
particular, the limiting cases 
$R_{\perp}$ = 1 for soft or collinear primary gluon
emission and the proper $gg$ limit for the parallel $q \bar q$
($\Theta_{+-}$ = 0) configuration, as expected.

The role of the large-$N_C$ limit 
can be investigated also by studying  the production rate in
3-jet events normalized 
to the sum of rates from the corresponding 2-jet events (dipoles) 
with opening angle
$\Theta_{1+}$ and $\Theta_{1-}$ respectively:
\ba
\label{rtilde} 
\tilde R_{\perp} &\equiv& \frac{dN_{\perp}^{q\bar q g}}{dN_{\perp}^{q\bar
q}(\Theta_{1+}) + dN_{\perp}^{q\bar q}(\Theta_{1-})} \\ 
&=& \frac{N_C^2}{N_C^2 -1} \Bigl( 1 - \frac{1}{N_C^2} \frac{1 - \cos
(\Theta_{1+} + \Theta_{1-} )}{2 - \cos \Theta_{1+} - \cos \Theta_{1-}} \Bigr)
\nonumber 
\ea
This ratio measures directly the deviation from the large-$N_C$ limit 
$\tilde R_{\perp}$= 1 and 
thereby from the $q\bar q$-dipole approximation. This approximation is not
necessarily limited towards soft particle production. For the simple case of
Mercedes-like events ($\Theta_{1+} = \Theta_{1-} = \Theta_{+-}$) one obtains 
$\tilde R_{\perp} = 17/16 = 1.06$. 
The 2-jet rates for relative angle $\Theta_{ij}$ which appear in the
denominator of eq.~\eref{rtilde} could be found experimentally from the
corresponding $q\bar q \gamma$ final states.

\begin{table}     
 \begin{center}
 \vspace{4mm}
 \begin{tabular}{||c|c|c||}
  \hline
 & $R_{\perp}$ & $R_{\perp}$ (large $N_C$) \\ 
  \hline
 $\Theta_{1+} = \pi - \Theta_{1-}$ & 1 & 1 \\ 
 (collinear or soft gluons) & & \\ 
 $\Theta_{1+} =  \Theta_{+-} = \frac{5}{6} \pi$ & 
1.21 &  
1.18 \\ 
 $\Theta_{1+} = \Theta_{+-} = \frac{3}{4} \pi$ & 
 1.42 & 
 1.35 \\ 
 $\Theta_{1+} = \Theta_{1-} = \frac{2}{3} \pi$ & 
 1.59  &  
  1.5 \\ 
 (Mercedes) & & \\ 
 $\Theta_{1+} = \Theta_{1-} = \pi$ & $\frac{N_C}{C_F}$ = 2.25 & 2 \\ 
 ($q\bar q$ antiparallel to $g$) & & \\ 
 \hline 
 \end{tabular}
 \end{center}
\caption{Prediction for the ratio $R_{\perp} = dN_{\perp}^{q\bar q g}/dN^{q\bar
q}_{\perp}$ from \protect\eref{rperp} 
and its large-$N_C$-approximation~\eref{rperpln} for different configurations
of the $q\bar q g$ events ($\Theta_{1+} \equiv \Theta_{gq}$, $\Theta_{1-} \equiv
\Theta_{g\bar q}$, 
$\Theta_{+-} = 2 \pi - \Theta_{1+} - \Theta_{1-}$).}
\label{tableperp}
\end{table}

Let us list a few further results:

\noindent {\it a)} 
A particularly simple situation is met for Mercedes-type events
where no jet identification is necessary for the above measurements. 

\noindent {\it b)} 
The large angle radiation is independent of the mass of the quark (for
$\Theta \gg m_Q/ E_{jet}$).

\noindent {\it c)} 
The above predictions for the ratios $R_{\perp},\tilde R_{\perp}$ are
derived for the soft particles according to the bremsstrahlung
formula~\eref{wqq}. One may also consider the particle
flow integrated over momentum, as in the discussion of the string/drag effect. In this case one has to
include all higher order contributions which take into account the fact that
the soft gluon is part of a jet generated from a primary parton. Then the
angular flow $dN/d\Omega_{\vec{n}}$ is given by the 
product of the radiation factor 
$W(\vec{n})$ and a ``cascading factor''\cite{adkt}. For the ratio of multiplicity flows
one obtains the same predictions~\eref{rperp},\eref{rtilde} 
as the cascading factors cancel.
It will be interesting to see to what extent the predicted angular dependence
for both quantities -- the multiplicity flows and the soft particle yields --
are satisfied experimentally. 

\noindent {\it d)} 
The similarity of particle flows in 3-jet and 2-jet events with corresponding
angles $\Theta_{1+}$, $\Theta_{1-}$, as expected in the large-$N_C$
approximation, should also apply to further details of the final state such as
the particle ratios. Since the Lorentz transformation along the boost direction
produces a larger  drag for heavier particles, one may expect 
that after the boost the $K/\pi$ and $p/\pi$ ratios for soft
particles decrease and then the same is true for soft particles in 3-jet
events in comparison to 2-jet events in their rest frame\footnote{This
expectation has been verified in the JETSET Monte Carlo\cite{jetset}. 
We thank T. Sj\"ostrand for
providing us with this information.}.


\section{Extension to different reactions}

We consider here processes with primary hadrons or photons with dominantly
2-jet final states. This includes semihard processes which are initiated by 
partonic 2-body scatterings. In case of quark exchange the two outgoing jets
originate from colour triplet charges and $I_0$ should be as in $e^+ e^-$
annihilation; in case of gluon exchange $I_0$ should be 
about twice as large ($N_C/C_F$) as in a $gg$ jet system.
In 2-jet events the limiting density $I_0$ may be most conveniently
determined as limit of $dn/dyd^2p_T$ at $y\approx 0$ for $p_T\to
0$.
Final states with several well separated jets can be
treated in analogy to the $e^+ e^- \to $ 3-jet events discussed before.

\medskip
\noindent{\it (a) quark exchange processes}\\
A simple example is the process $\gamma \gamma \to q \bar q \to$ 2-jets
which dominates if either the virtuality $Q^2$ of the initial photon or
the scattering angle photon-jet is sufficiently large (see, e.g.
\cite{kz}).
In this case the soft production intensity $I_0$ should be the same as in
$e^+ e^-$ annihilation.

Another example is deep inelastic scattering at large $Q^2$. In this case
the current fragments in the Breit frame  
are expected to have the same
characteristics as the quark fragments in one hemisphere of
 $e^+ e^-$ annihilation (for a QCD analysis, see\cite{gdkt}),
and this is indeed observed for not too small $Q^2$\cite{h1,zeus}.
Therefore one expects again the same limit $I_0$ as in
$e^+ e^-$ annihilation.

\medskip
\noindent{\it (b) gluon exchange processes}\\
These are expected to dominate for virtual $\gamma_V \gamma$
or $\gamma_V p$ scattering ($Q^2 \gsim$ few GeV$^2$)  
through the photon gluon fusion subprocess
$\gamma g \to q \bar q$ at small Bjorken $x$.
In this case (for a fast  $q \bar q$ pair in one 
hemisphere with small opening angle) the 
 particle jets emerging in opposite directions 
 are generated by an effective colour octet
emitter and the  soft intensity $I_0$ should be about twice as large
 as in $e^+ e^-$ annihilation.

Another example is hadron-hadron collision with a particle or jet at
moderate $p_T$ ($\gsim$ 1--2 GeV) at small angles so that the overall 2-jet
structure is maintained. This type of scattering is dominated by gluon
exchange and the same increase of $I_0$ is therefore expected.

\medskip
\noindent{\it (c) soft collisions (minimum bias events)}\\
These processes (with initial hadrons or real photons) are not so well
understood theoretically
as the hard ones but it might be plausible to extrapolate the gluon
exchange process towards small $p_T$\cite{ln}.
Experimental data at ISR energies, however, do not follow the expectation
of a doubling of $I_0$ in comparison to $e^+ e^-$ annihilation, rather the
intensities are found to be similar\cite{ppscal}. 
On the other hand, $I_0$ roughly doubles when going from $\sqrt{s} \sim$ 20 GeV
\cite{na22,ISR} to $\sqrt{s}$ = 900~GeV  at the collider\cite{coll}.
If additional incoherent sources like weak decays can be excluded, such behaviour
could indicate the growing importance of one-gluon exchange expected from the
perturbative picture. Then a saturation at $I_0^{hh}/I_0^{e^+e^-} \sim 2$
is expected and no additional increase for a semihard $p_T$ trigger.
A rise of $I_0^{hh}$ could also result from incoherent multiple
collisions of partons (e.g. \cite{pythia})  
which recently has also been postulated
for $\gamma p$ collisions\cite{sz}.

It should be noted that events from all three classes $a$, $b$ and $c$ can
occur in the same reaction depending on the external kinematical constraints
(triggers). For example, if $Q^2$ for fixed hadronic energy $W$ 
in $\gamma p$ or $\gamma\gamma$ collisions 
is decreased from large $Q^2\sim W^2$ towards zero 
one changes from (a) to (b) and
finally to (c). Also, a dependence on the reference frame is expected,
case (a) in Breit frame, case (b) typically in $cms$ frame for small Bjorken
$x$. 

\medskip
\noindent{\it Rapidity dependence of $I_0$.}\\
The frame dependence can be studied conveniently in these exchange
processes by considering $I_0(y)= \frac{dn}{dy d^2p_T} 
\bigr|_{p_T \to 0}$
as a function of rapidity $y$, measured, say, in the $cms$ frame. 
The Breit frame is reached by a Lorentz transformation in the direction of the
incoming photon, and corresponds to a different rapidity, $y_{Breit}$, in the
$cms$. According to the above discussion, one would then expect
for the processes with virtual photons at large $Q^2$ 
in general a ``quark plateau" of
$I_0(y)$ in the current region of length $\Delta y \approx \log (Q/m)$
near $y_{Breit}$ and a transition to a 
``gluon plateau" in the complementary region of length 
$\Delta y \approx \log (W^2/Qm)$ where $m$ is a typical particle mass. The
existence of a plateau corresponds to the energy independence of $I_0$ as seen 
in $e^+e^-$ annihilation. So, if gluon exchange occurs with sufficient hardness
in the process, $I_0$ should develop a step like behaviour but never become
larger than twice the $I_0$ value in $e^+e^-$ annihilation. The height of the
``plateau'' is a direct indicator of the underlying exchange process (quark or
gluon like). 

\section{Conclusions}

The analytical perturbative approach to multiparticle production, based on the
Modified Leading Log Approximation and Local Parton Hadron Duality (LPHD)
has proven to be successful in the description of various inclusive characteristics
of jets. It is of importance to investigate the limitations of this picture, in
particular in the soft region where non-perturbative effects are expected to occur.

It is remarkable that the intensity of the soft hadron production follows to a good
approximation a scaling law in a range of two orders of magnitude in the $cms$ energy
of $e^+ e^-$ annihilation (1.6-140 GeV). Such a scaling law is derived analytically for
the soft gluons in the jet and follows directly from the coherence of the soft gluon
radiation from all emitters. It appears  that the production of hadrons which is known
to proceed through many resonance channels nevertheless can be simulated in the average
through a parton cascade down to a small scale of a few 100 MeV as suggested by LPHD.
The scaling behaviour down to the very low $cms$ energy of 1.6 GeV can only
be explained within a parton model description if the cut-off scale $Q_0$
is well below 1 GeV.

It will be interesting to investigate the suggested scaling property~(1) 
and its possible violation further in the same experiment 
to avoid systematic
effects; this seems to be feasible at LEP and HERA for the simplest processes.
In particular, the effect of weak decays on scaling violations should be
clarified.
The sensitivity of the soft particle production to the effective colour 
charge of the primary emitters can be tested through the transverse 
production rates in multijet events.
The soft radiation in  $e^+ e^-$ annihilation can be used as a 
standard scale in the comparison of various processes. 
In this way the soft particle production can be a sensitive probe
of the underlying partonic process and the contributions 
from incoherent sources.

\section*{Acknowledgements} 
We would like to thank Yu. L. Dokshitzer, P. Minkowski 
and T. Sj\"ostrand for useful
discussions. This work was supported in part by the UK PPARC
and the EU Training and Mobility
Program ``Hadronic Physics with High Energy Electromagnetic Probes", Network
FMRX-CT96-0008.


\begin{figure}[p]
          \begin{center}
\mbox{\epsfig{file=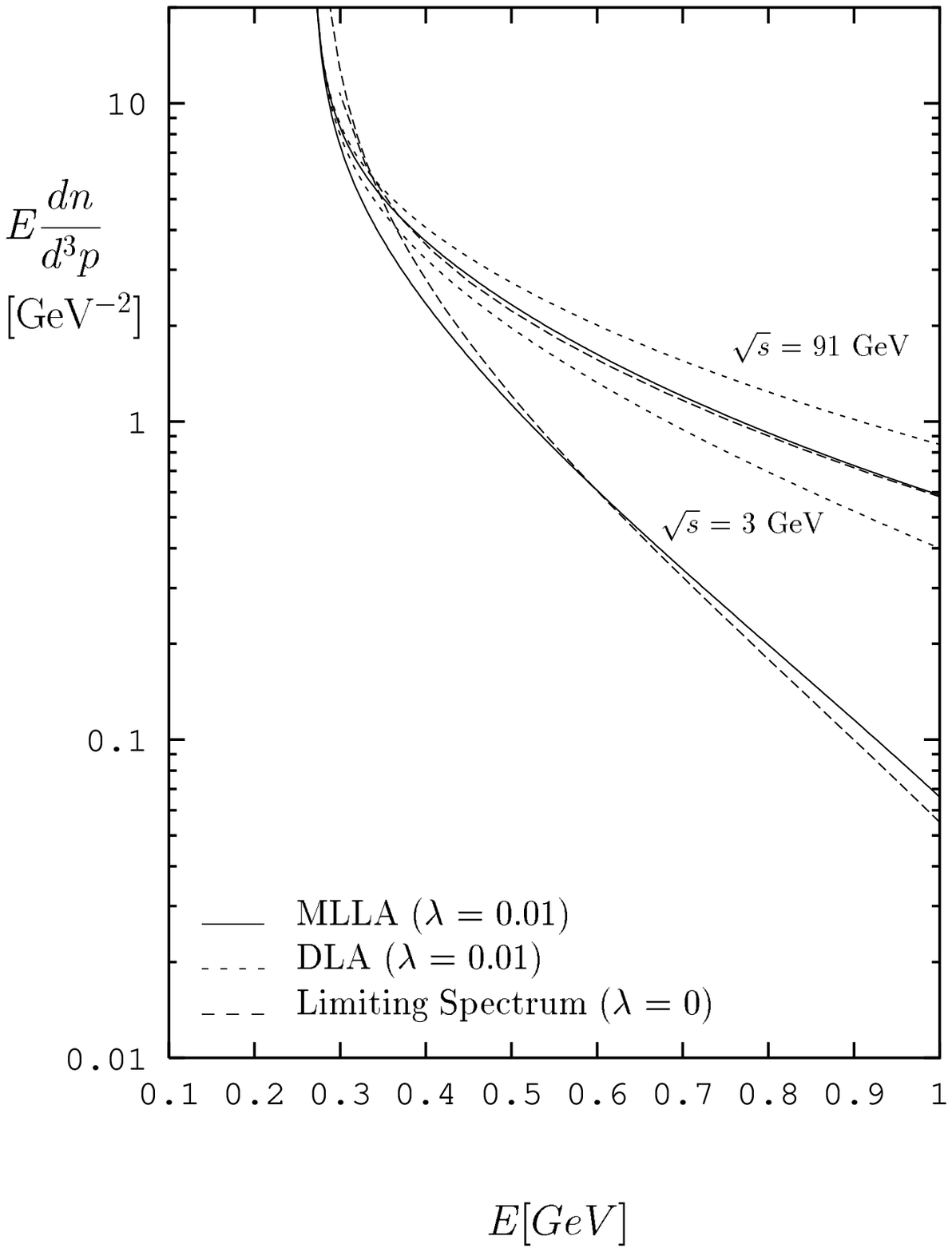,bbllx=4.5cm,bblly=9.5cm,bburx=16.5cm,bbury=26.cm}}
       \end{center}
\caption{Invariant density $E dn/d^3p$ 
as a function of the particle energy $E$ for $Q_0$ = 270 MeV. 
Predictions  at $cms$ energies of $\protect\sqrt{s}$
= 3 GeV (lower three curves) and 91 GeV (upper three curves)  
using $E dn/d^3p = 2\cdot 4/9 K_h D_g^g(\xi_E)/[4 \pi (E^2 - Q_0^2)]$ 
with $D_g^g$
computed in MLLA (eqs.~(\protect\ref{dmlla},\protect\ref{duetermini}), 
DLA (eq.~\protect\eref{duetermini}) 
and the Limiting  Spectrum (normalization $K_h$ = 0.45, 0.45 and 1.125
respectively).} 
\label{fixed}
\end{figure}


\begin{figure}[p]
          \begin{center}
\mbox{\epsfig{file=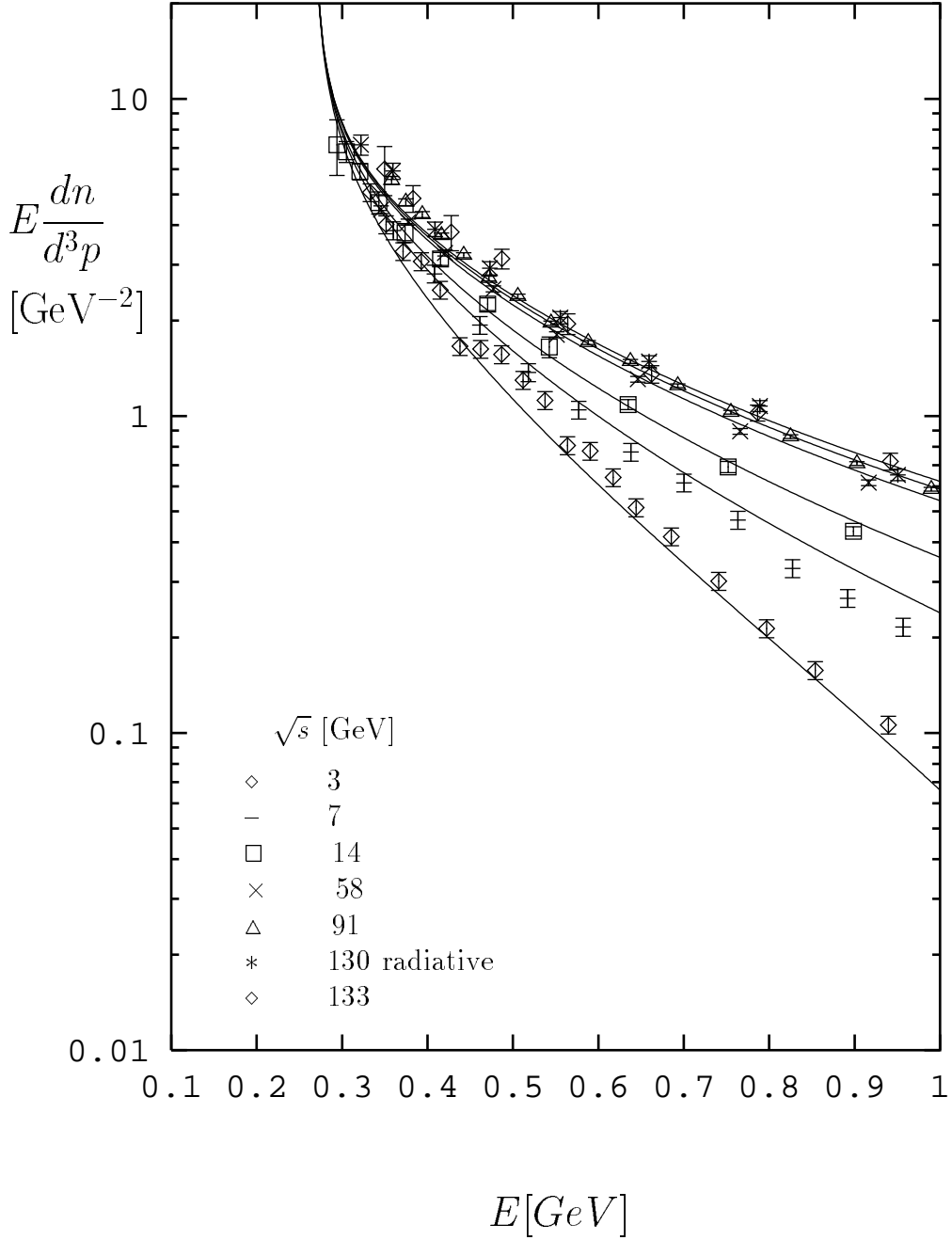,bbllx=4.5cm,bblly=9.5cm,bburx=16.5cm,bbury=26.cm}}
       \end{center}
\caption{Invariant density $E dn/d^3p$ of charged particles
in $e^+e^-$ annihilation 
as a function of the particle energy 
$E=\protect\sqrt{p^2+Q_0^2}$ at $Q_0$ = 270 MeV. 
Data points at various $cms$ energies
from  SLAC, TASSO and TOPAZ Collaborations, LEP-1 and
LEP-1.5\protect\cite{data,tasso,opal1} are compared to MLLA predictions 
(normalization as in Fig.~1 with 
$\lambda$ = 0.01, $K_h$ = 0.45).}
\label{chargedall}
\end{figure}


\begin{figure}[p]
\vfill \begin{minipage}{.45\linewidth}
          \begin{center}
\mbox{\epsfig{file=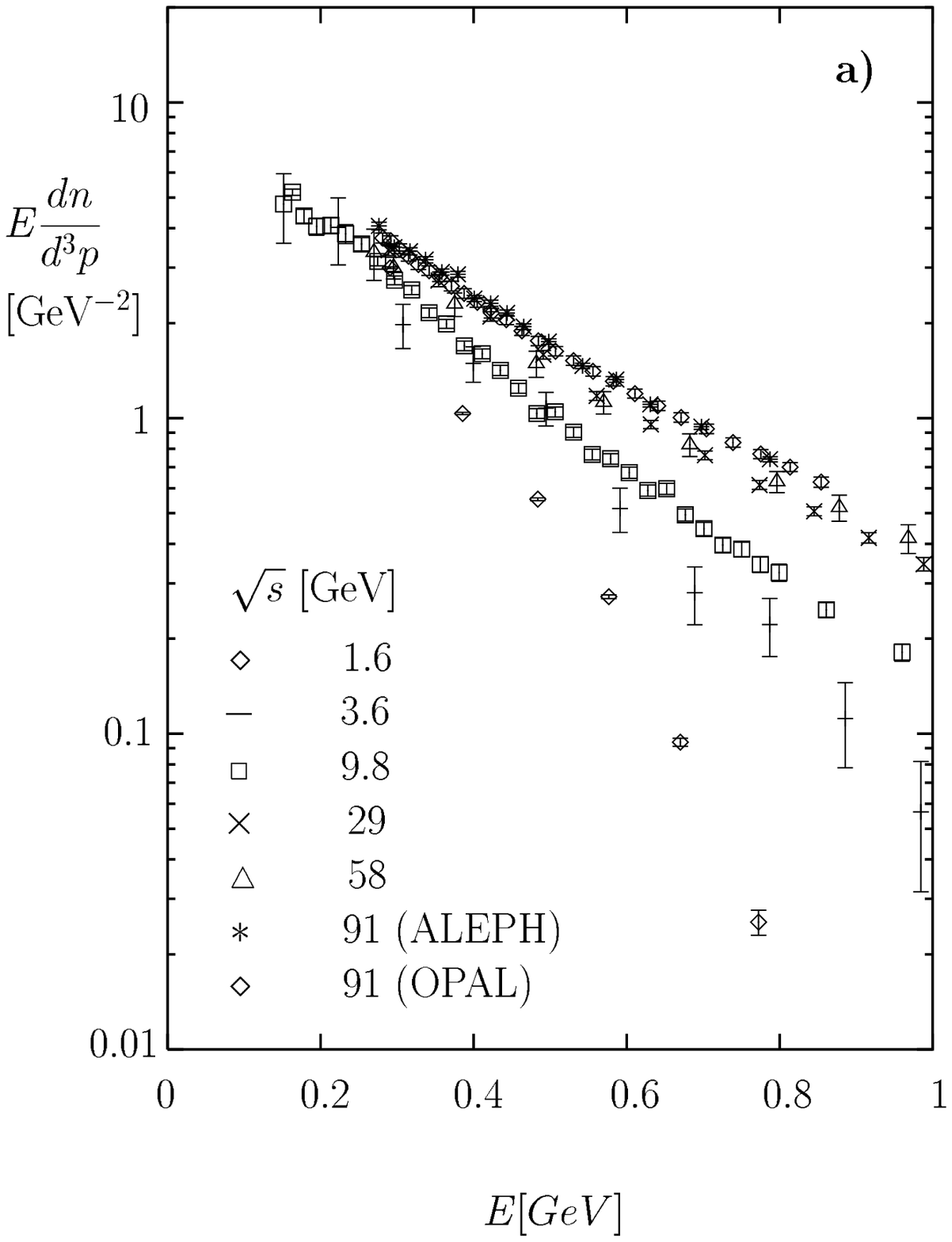,width=.6\linewidth,bbllx=5.5cm,bblly=9cm,bburx=13.cm,bbury=20.cm}}
          \end{center}
      \end{minipage}\hfill
      \begin{minipage}{.45\linewidth}
          \begin{center}
\mbox{\epsfig{file=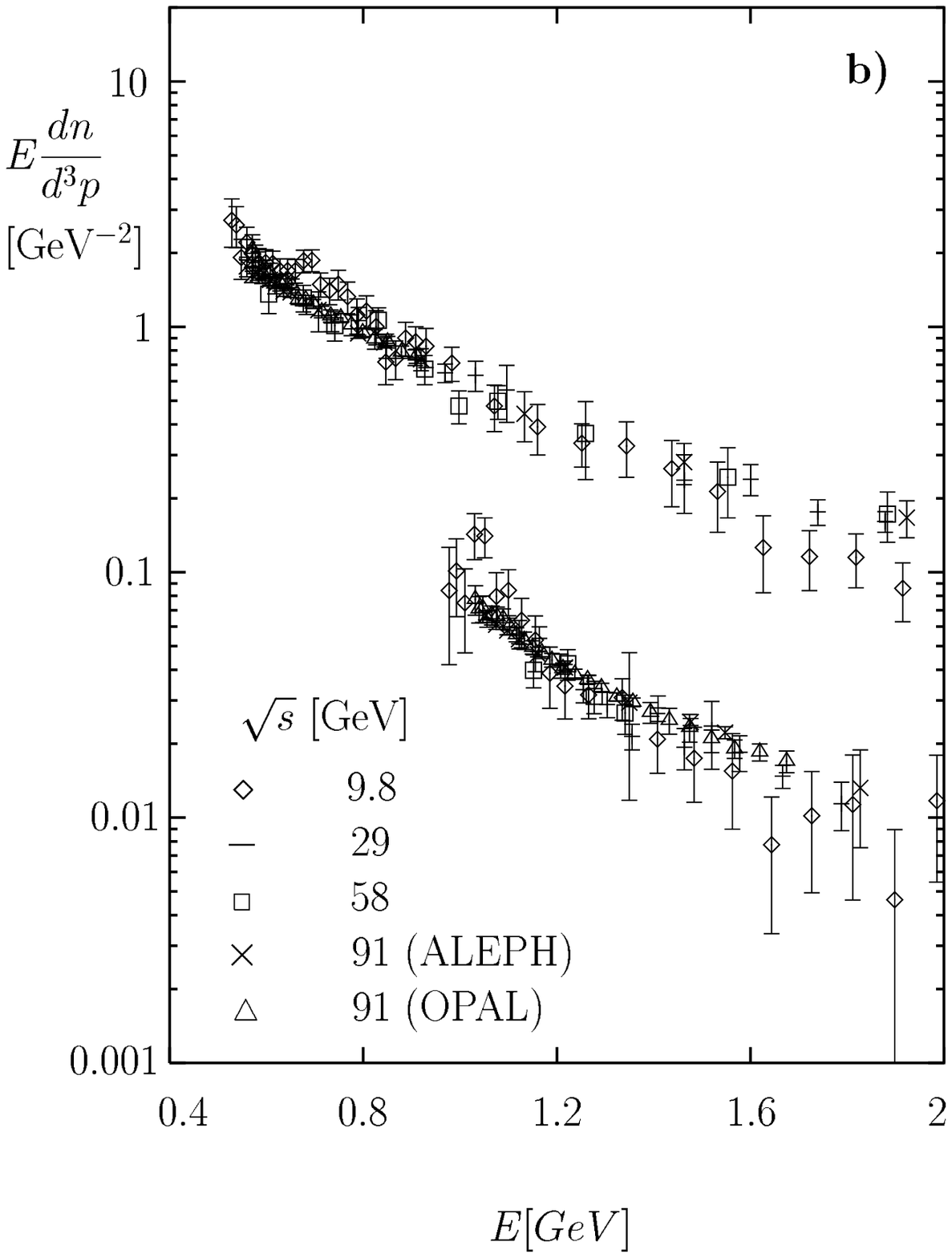,width=.6\linewidth,bbllx=5.5cm,bblly=9cm,bburx=13.cm,bbury=20.cm}}
          \end{center}
      \end{minipage}
\caption{{\bf a)} Invariant density $E dn/d^3p$ of charged pions in $e^+e^-$
annihilation 
as a function of the particle energy $E$ at $Q_0$ = 138 MeV at 
various $cms$ energies;
{\bf b)} the same as in {\bf a)}, but for charged kaons (with 
$Q_0$ = 494 MeV) and protons (with $Q_0$ = 938 MeV)   
\protect\cite{datapion,data}. The kaon 
distribution is multiplied by a
factor 10 for the sake of clarity.} 
\end{figure}

\end{document}